\documentclass[12pt]{article}
\pdfoutput=1
\usepackage[a4paper,text={16.8cm,22.4cm}]{geometry}
\usepackage{amsmath,amsfonts,braket,slashed,amssymb,tikz,bm,psfrag,graphicx,color,dsfont}
\usepackage{multicol}
\usepackage{ctable}
\usepackage[small,labelfont=bf]{caption}
\usepackage{hyperref}
\usepackage{subcaption}
\usepackage{cleveref}
\usepackage{bbm}
\usepackage{braket,tikz,psfrag,euscript}

\newcommand{\J}{{\EuScript J}}
\newcommand{\nsl}{\rlap{\hspace{0.25mm}/}{n}}
\newcommand{\nbsl}{\rlap{\hspace{0.25mm}/}{\bar n}}
\newcommand{\Dsl}{\rlap{\hspace{0.75mm}/}{D}}
\newcommand{\spac}{{\hspace{0.3mm}}}

\numberwithin{equation}{section}

\RequirePackage[sort&compress,square,comma,numbers]{natbib}
\allowdisplaybreaks
\begin{document}
\begin{titlepage}
\begin{flushright}
MITP/21-065\\ 
November 30, 2021
\end{flushright}
		
\vspace{0.5cm}
\begin{center}
\Large\bf\boldmath
Radiative Quark Jet Function with an External Gluon
\end{center}
		
\vspace{0.5cm}
\begin{center}
Ze Long Liu$^a$, Matthias Neubert$^{b,c}$, Marvin Schnubel$^{b}$ and Xing Wang$^{b}$\\[7mm]
{\sl ${}^a$Institut f\"ur Theoretische Physik \& Albert Einstein Center\\ 
Universit\"at Bern, Sidlerstrasse 5, CH-3012 Bern, Switzerland\\[2mm]
${}^b$PRISMA$+$ Cluster of Excellence \& Mainz Institute for Theoretical Physics\\
Johannes Gutenberg University, 55099 Mainz, Germany\\[2mm]
${}^c$Department of Physics \& LEPP, Cornell University, Ithaca, NY 14853, U.S.A.}
\end{center}
		
\vspace{0.8cm}
\begin{abstract}
Factorization theorems in soft-collinear effective theory at subleading order in power counting involve ``radiative jet functions'', defined in terms of matrix elements of collinear fields with a soft momentum emitted from inside the jet. Of particular importance are the radiative quark jet functions with an external photon or gluon, which arise e.g.\ in the factorization theorems for the Higgs-boson amplitudes $h\to\gamma\gamma$, $h\to gg$ and $gg\to h$ induced by light-quark loops. While the photon case has been studied extensively in previous work, we present here a detailed study of the radiative jet function with an external gluon. We calculate this jet function at one- and two-loop order, derive its one-loop anomalous dimension and study its renormalization-group evolution.
\end{abstract}
\end{titlepage}

\section{Introduction}

Soft-collinear effective theory (SCET) is a helpful tool to study high-energy cross sections and decay rates sensitive to several hierarchical mass scales (see e.g.\ \cite{Bauer:2002nz} and references therein). Deriving factorization theorems, consisting of functions each depending on a single scale, is an important step towards understanding the dynamics of such objects. Typically, the various functions receive contributions from different momentum regions in Feynman diagrams. The hard functions are the Wilson coefficients obtained by matching the Standard Model onto SCET. Jet and soft functions, on the other hand, are defined in terms of matrix elements of collinear or soft fields in the effective theory, respectively. In particular, the jet functions consist of non-local products of collinear fields dressed by Wilson lines, and they often depend on an intermediate scale between the hard-scattering scale of the process and the soft scale.

In the past years, there has been a growing interest in understanding factorization at subleading power in scale ratios. Beyond the leading power a large variety of hard, collinear and soft functions appear. In particular, while at leading power soft emissions are eikonal and can be described by soft Wilson lines, at subleading power the emission of soft quarks needs to be taken into account. At subleading order in the SCET expansion, there is a unique interaction that couples a soft quark to collinear quarks and gauge fields. In the notation of \cite{Beneke:2002ph} it reads 
\begin{equation}\label{subleadingL}
   {\cal L}_{q\,\xi_n}^{(1/2)}(x)
   = \bar q_s(x_-)\,W_n^\dagger(x)\,i\Dsl_n^\perp\,\xi_n(x) + \mbox{h.c.} \,, 
\end{equation}
where $\xi_n$ is a collinear quark spinor subject to the constraint $\nsl\,\xi_n=0$. It is related to the conventional QCD quark field by $\xi_n=\frac14\spac\nsl\nbsl\,\psi$, where $n^\mu=(1,\bm{n})$ is a null-vector aligned with the jet direction, and $\bar n^\mu=(1,-\bm{n})$ is a null-vector pointing in the opposite direction, such that $n\cdot\bar n=2$. The quantity $W_n$ denotes a collinear Wilson line (a path-ordered exponential of a line integral over the collinear gluon field $\bar n\cdot A_n$), $D_n^\mu$ is a covariant collinear derivative, and $q_s$ is a soft quark field. The collinear particles carry large momentum along the direction $n^\mu$, whereas the momenta of soft fields are small. As a consequence, in interactions with collinear fields a soft quark field must be multipole expanded \cite{Beneke:2002ph}, and we denote $x_-^\mu=(\bar n\cdot x)\,\frac{n^\mu}{2}$ in the above relation.

Scale factorization at subleading power in SCET turns out to be an unexpectedly rich and complicated problem due to the presence of endpoint divergences in convolution integrals \cite{Moult:2019mog,Beneke:2019kgv,Moult:2019uhz,Beneke:2019oqx,Moult:2019vou,Liu:2019oav,Beneke:2020ibj}. Only recently, the derivation of a renormalized factorization theorem for a subleading-power quantity, in which endpoint divergences are regularized using plus-type subtractions, has been accomplished for the first time \cite{Liu:2019oav,Liu:2020tzd,Liu:2020wbn}. The pseudo-observable considered in these works was the Higgs-boson decay $h\to\gamma\gamma$ induced via a $b$-quark loop. In the factorization formula, the radiative quark jet function with an external collinear photon arose \cite{Liu:2020ydl}, which is an object well known from earlier studies of QCD factorization in the radiative $B$-meson decay $B^-\to\gamma\ell^-\bar\nu$ \cite{Lunghi:2002ju,Bosch:2003fc}. The term ``radiative jet function'' indicates that, in contrast to the jet functions appearing in leading-power SCET problems, a soft quark is emitted from inside the jet \cite{DelDuca:1990gz,Bonocore:2015esa,Bonocore:2016awd,Moult:2019mog}. 

Generalizing the $h\to\gamma\gamma$ decay to its non-abelian cousins, i.e.\ the decay $h\to gg$ or the production process $gg\to h$, one encounters the radiative quark jet function $J_g(p^2)$ with an external collinear gluon \cite{HiggsGluGlu}, which is the main subject of this work. In analogy with the treatment of the radiative jet function with an external photon in \cite{Liu:2020ydl}, we define the function $J_g(p^2)$ in terms of the matrix element 
\begin{equation}\label{defJg}
   \int d^dx\,e^{ip_s\cdot x_-}\,\langle g(k,a)|\,
    T\,\big( W_n^\dagger\,i\Dsl_n^\perp\,\xi_n \big)(x)\,\big( \bar\xi_n W_n \big)(0)\,|0\rangle 
   = g_s\spac t_a\,\rlap/\varepsilon_\perp^*(k)\,\frac{\nsl}{2}\,\frac{i\bar n\cdot k}{p^2+i0}\,J_g(p^2) \,, 
\end{equation}
where $g_s$ is the strong coupling, $k$, $\varepsilon$ and $a$ are the momentum, polarization vector and color index of the external gluon, respectively, $p_s$ denotes the momentum carried away by the soft quark in (\ref{subleadingL}), and $p\equiv k+p_{s+}$. Note that $p_s\cdot x_-=p_{s+}\cdot x$ with $p_{s+}^\mu=(n\cdot p_s)\,\frac{\bar n^\mu}{2}$. Because of the conditions $k^2=0$ and $p_s^2=0$, the jet function $J_g(p^2)$ depends on the single non-trivial kinematic invariant 
\begin{equation}
   p^2\equiv (k+p_{s+})^2 = \bar n\cdot k\,n\cdot p_s \,.
\end{equation}
The function $J_g(p^2)$ plays a central role in deriving exact $d$-dimensional refactorization conditions, which are at the heart of our method for removing endpoint divergences by means of plus-type subtractions \cite{Liu:2019oav,Liu:2020tzd,Liu:2020wbn}. The study of this radiative jet function thus offers a new window to investigate the structure of power-suppressed corrections to processes with external collinear gluons.

\section{Two-loop calculation of the bare jet function}

We write the result for the bare jet function in $d=4-2\epsilon$ spacetime dimensions in the general form
\begin{equation}\label{Jgbare}
   J_g^{(0)}(p^2) = 1 + \frac{\alpha_{s,0}}{4\pi} \left( -p^2-i0 \right)^{-\epsilon} J_{g,1}^{(0)}
    + \left( \frac{\alpha_{s,0}}{4\pi} \right)^2 \left( -p^2-i0 \right)^{-2\epsilon} J_{g,2}^{(0)}
    + \mathcal{O}(\alpha_s^3) \,.
\end{equation}
In our calculations we adopt the background field method for the external gluon \cite{tHooft:1975uxh,Boulware:1980av,Abbott:1980hw,Abbott:1981ke,Meissner:1986tr}. It is most convenient to perform the calculation of the expansion coefficients $J_{g,1}^{(0)}$ and $J_{g,2}^{(0)}$ in light-cone gauge $\bar n\cdot A_n=0$, such that the collinear Wilson lines $W_n=1$ are trivial. The smaller number of Feynman diagrams and the absence of ghost contributions allow for an efficient computation in this gauge. The free gluon propagator with momentum $l$ in light-cone gauge is given by 
\begin{equation}
   \frac{i}{l^2+i0} \left( - g^{\mu\nu} + \frac{{\bar n}^\mu l^\nu + {\bar n}^\nu l^\mu}{\bar n\cdot l} \right) ,
\end{equation}
where we do not adopt the Mandelstam--Leibbrandt prescription to regularize the singularity at ${\bar n}\cdot l=0$ (see \cite{Becher:2010pd} for more details). As a cross check of our results, we have also performed the calculation in a general covariant gauge, finding identical results. 

\begin{figure}[t]
\begin{center}
\includegraphics[width=0.9\textwidth]{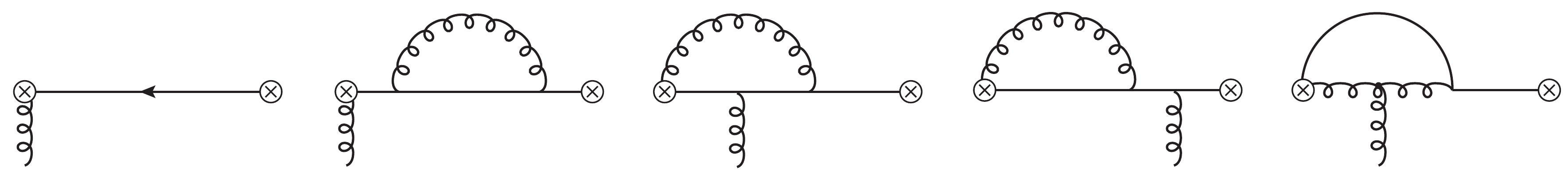}
\caption{\label{fig:diagrams1loop} 
Tree-level and one-loop diagrams contributing to the jet function in light-cone gauge. The left cross corresponds to the spacetime point $x$, where the soft momentum $p_s$ flows out, see (\ref{defJg}). The right cross corresponds to the point 0. The fourth diagram is scaleless and evaluates to zero.}
\end{center}
\end{figure}

The diagrams contributing at one-loop order to the jet function are shown in Figure~\ref{fig:diagrams1loop}. Evaluating these graphs we obtain 
\begin{equation}
   J_{g,1}^{(0)}
   = (C_F-C_A)\,e^{\epsilon\gamma_E}\,\frac{\Gamma(1+\epsilon)\,\Gamma^2(-\epsilon)}{\Gamma(2-2\epsilon)}\,
    (2-4\epsilon-\epsilon^2) \,,
\end{equation}
where $C_F=(N_c^2-1)/(2 N_c)$ and $C_A=N_c$ are the eigenvalues of the quadratic Casimir operators in the fundamental and adjoint representations of $SU(N_c)$. Remarkably, to all orders in $\epsilon$ this result differs from the corresponding expression for the jet function with an external photon only in the color factor, which is $C_F$ in the photon case. We will see below that such a simple relation no longer holds beyond one-loop order.

\begin{figure}[t]
\begin{center}
\includegraphics[width=\textwidth]{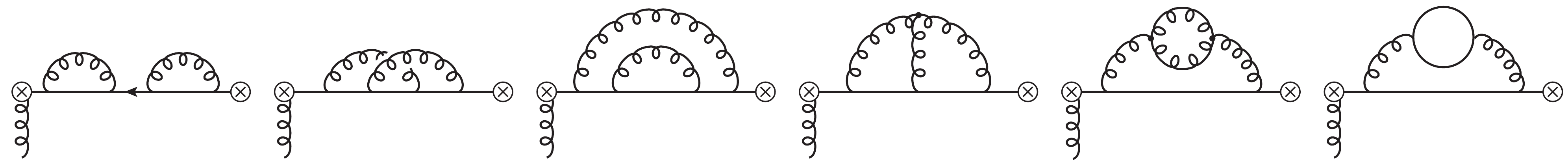}\\[3mm]
\includegraphics[width=\textwidth]{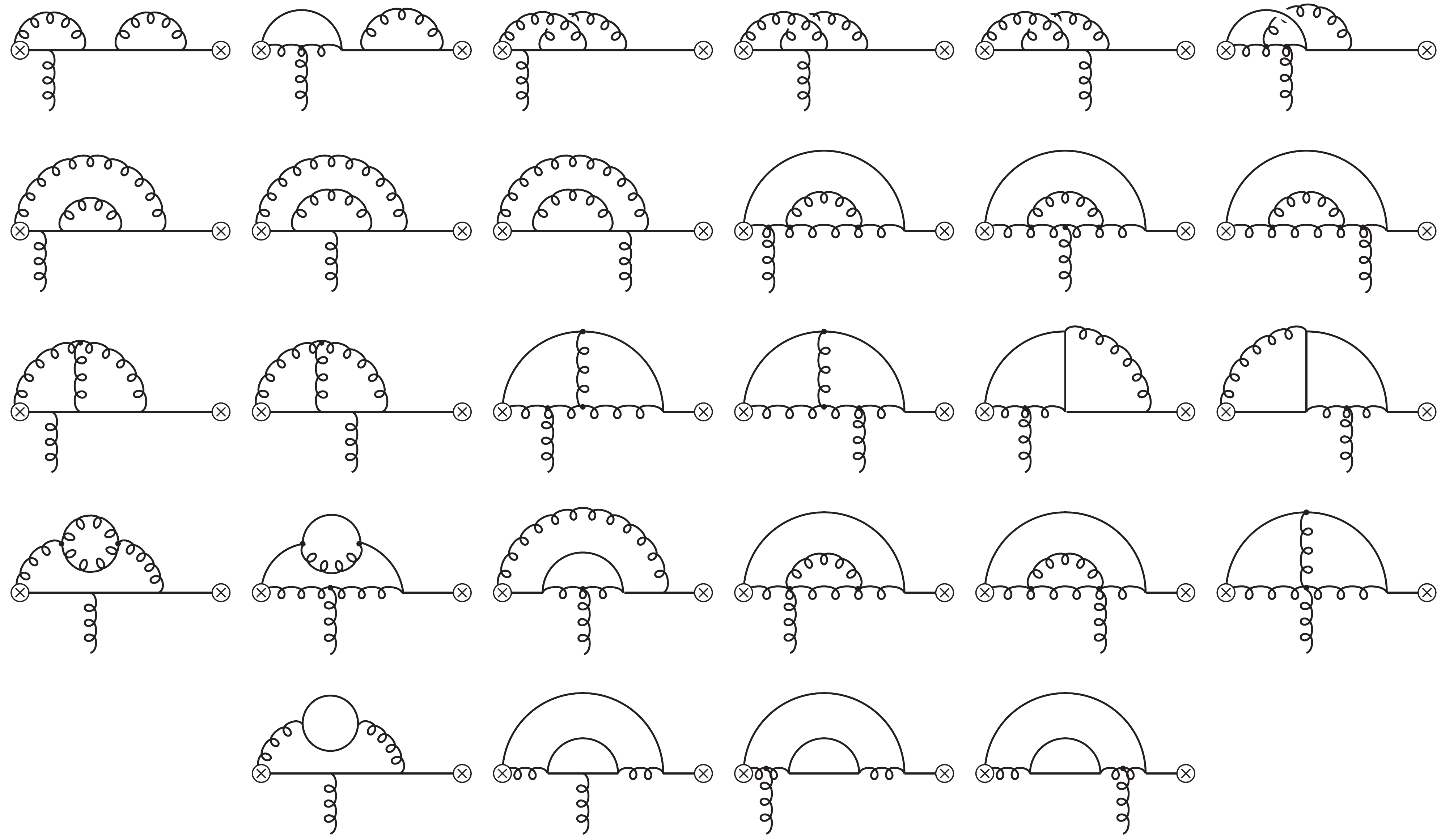}
\caption{\label{fig:diagrams2loop}
Two-loop diagrams contributing to the jet function in light-cone gauge.}
\end{center}
\end{figure}

At two-loop order, the diagrams contributing in light-cone gauge are presented in Figure~\ref{fig:diagrams2loop}. Using simplifications of the Dirac and Lorentz structures, each diagram can be expressed as a linear combination of scalar two-loop integrals of the form
\begin{equation}
   \int d^dl_1 \int d^d l_2\,\frac{1}{\prod_{i=1}^{15} \mathcal{D}_i^{a_i}} \,,
\end{equation}
where the propagators are (omitting the $-i0$ prescription)
\begin{equation}
\begin{aligned}
   \mathcal{D}_1 &=-l_1^2 \,, & \mathcal{D}_2 &=-l_2^2 \,, & \mathcal{D}_3 &=-(l_1+l_2)^2 \,, \\
   \mathcal{D}_4 &=-(l_1+p)^2 \,, & \mathcal{D}_5 &=-(l_2+p)^2 \,, & \mathcal{D}_6 &=-(l_1+l_2+p)^2 \,, \\
   \mathcal{D}_7 &=-(l_1+p-k)^2 \,, & \mathcal{D}_8 &=-(l_2+p-k)^2 \,, & \mathcal{D}_9 &=-(l_1+l_2+p-k)^2 \,, \\ 
   \mathcal{D}_{10} &=-\bar{n}\cdot l_1 \,, & \mathcal{D}_{11} &=-\bar{n}\cdot l_2 \,,
    & \mathcal{D}_{12} &=-\bar{n}\cdot (l_1+l_2) \,, \\ 
   \mathcal{D}_{13} &=-\bar{n}\cdot (l_1+p) \,, & \mathcal{D}_{14} &=-\bar{n}\cdot ( l_2+p) \,,
    & \mathcal{D}_{15} &=-\bar{n}\cdot (l_1+l_2+p) \,.
\end{aligned}
\end{equation}
After performing partial-fraction decompositions, all scalar Feynman integrals are mapped onto linear complete integral families. We then use the integration-by-parts method as implemented in {\verb+LiteRed+} \cite{Lee:2013mka}, {\verb+FIRE6+} \cite{Smirnov:2019qkx} and {\verb+Kira+} \cite{Klappert:2020nbg} to reduce these Feynman integrals to linear combinations of master integrals. With the help of algebraic relations, we have been able to express all master integrals in terms of integrals encountered in the calculation of the jet function with an external photon \cite{Liu:2020ydl}. In this way, we obtain
\begin{equation}\label{eq:2.6}
   J_{g,2}^{(0)} = C_F^2\spac K_{FF} + C_F\spac C_A\spac K_{FA} + C_A^2\spac K_{AA} 
    + C_F\spac T_F\spac n_f\spac K_{F\spac n_f}+ C_A\spac T_F\spac n_f\spac K_{A\spac n_f} \,,
\end{equation}
with
\begin{align}\label{eq:2.7}
   K_{FF} &= \frac{2}{\epsilon^4} - \frac{1}{\epsilon^2} \left( 2+\frac{\pi^2}{3} \right)
    - \frac{1}{\epsilon} \left( 4+\frac{\pi^2}{2}+\frac{46\zeta_3}{3} \right) 
    - \frac{13}{2} - \frac{\pi^2}{6}-39\zeta_3+\frac{\pi^4}{5} + {\cal O}(\epsilon) \,, \notag\\
   K_{FA} &= -\frac{4}{\epsilon^4} + \frac{11}{6\epsilon^3} 
    + \frac{1}{\epsilon^2} \left( \frac{139}{18}+\frac{\pi^2}{2} \right)
    + \frac{1}{\epsilon} \left( \frac{319}{27}-\frac{\pi^2}{18}+\frac{80\zeta_3}{3} \right) \notag\\
   &\quad + \frac{1087}{162} - \frac{83\pi^2}{54} + \frac{485\zeta_3}{18} - \frac{49\pi^4}{360} 
    + {\cal O}(\epsilon) \,, \notag\\
   K_{AA} &= \frac{2}{\epsilon^4} - \frac{11}{6\epsilon^3} 
    - \frac{1}{\epsilon^2} \left( \frac{103}{18}+\frac{\pi^2}{6} \right) 
    - \frac{1}{\epsilon} \left( \frac{413}{54}-\frac{11\pi^2}{18}+\frac{34\zeta_3}{3} \right) \\
   &\quad + \frac{100}{81} + \frac{47\pi^2}{27} + \frac{259\zeta_3}{18} - \frac{23\pi^4}{360} 
    + {\cal O}(\epsilon) \,, \notag\\
   K_{F\spac n_f} &= - \frac{2}{3\epsilon^3} - \frac{10}{9\epsilon^2} 
    - \frac{1}{\epsilon} \left( \frac{20}{27}-\frac{\pi^2}{9} \right) 
    + \frac{230}{81} + \frac{5\pi^2}{27} + \frac{64\zeta_3}{9} + {\cal O}(\epsilon) \,, \notag\\
   K_{A\spac n_f} &= \frac{2}{3\epsilon^3} + \frac{10}{9\epsilon^2}
    + \frac{1}{\epsilon} \left( \frac{11}{27}-\frac{2\pi^2}{9} \right)
    - \frac{491}{81} - \frac{10\pi^2}{27} - \frac{106\zeta_3}{9} + {\cal O}(\epsilon) \,, \notag
\end{align}
where $T_F=\frac12$, and $n_f$ denotes the number of active quark flavors. The coefficients $K_{FF}$ and $K_{F\spac n_f}$ are identical to the corresponding coefficients in the photon case derived in \cite{Liu:2020ydl}, but for the remaining coefficients no simple relation between the two jet functions can be found past one-loop order. In the Appendix we give the expansions of these coefficients to second order in the dimensional regulator $\epsilon$. These results would be needed to obtain the renormalized jet function at three-loop order.

\section{Renormalization of the jet function}

In the $\overline{\text{MS}}$ subtraction scheme, the radiative jet function $J_g(p^2)$ can be renormalized multiplicatively in the convolution sense. We write
\begin{equation}\label{eq:Renormalizationdef}
   J_g(p^2,\mu) = \int_0^\infty\!dx\,Z_{J_g}(p^2,x p^2;\mu)\,J_g^{(0)}(x p^2) 
    \qquad \text{(scheme\,1)} \,.
\end{equation}
Note that jet functions at different values of $p^2$ mix under renormalization. For the analogous case of the radiative jet function with an external photon studied in \cite{Liu:2020ydl}, this ``non-local'' renormalization condition was derived using the fact that this jet function appears in the QCD factorization theorem for the $B^-\to\gamma\ell^-\bar\nu$ decay amplitude \cite{Bosch:2004th} along with the leading-order light-cone distribution amplitude of the $B$ meson \cite{Grozin:1996pq}, whose renormalization properties have been derived long ago in \cite{Lange:2003ff}. In the present case, we can derive the one-loop expression for the renormalization factor $Z_{J_g}$ by exploiting the fact that the function $J_g(p^2)$ enters the factorization theorem for the Higgs-boson production amplitude in gluon-gluon fusion, $gg\to h$, which is structurally very similar to the factorization theorem for the $h\to\gamma\gamma$ decay amplitude derived in \cite{Liu:2019oav,Liu:2020tzd,Liu:2020wbn}. Using the renormalization-group (RG) equations for the relevant hard and soft functions entering this factorization theorem,\footnote{The relevant soft function can be obtained from the soft function in the $h\to\gamma\gamma$ process \cite{Liu:2020eqe} by an appropriate replacement of color factors in each Feynman diagram.} 
we find that the renormalized jet function in (\ref{eq:Renormalizationdef}) obeys the RG evolution equation \cite{HiggsGluGlu}
\begin{equation}\label{eq:rge}
   \frac{d}{d\ln\mu}\,J_g(p^2,\mu)
   = - \int_0^\infty\!dx\,\gamma_{J_g}(p^2,x p^2;\mu)\,J_g(x p^2,\mu) \,,
\end{equation}
where at one-loop order the anomalous dimension is given by
\begin{equation}\label{eq:gammaJg}
\begin{aligned}
   \gamma_{J_g}(p^2,x p^2;\mu)
   &= \left[ (C_F-C_A)\,\frac{\alpha_s}{\pi}\,\ln\frac{-p^2-i0}{\mu^2} 
    - \frac{\gamma_{g,0}'\,\alpha_s}{4\pi} \right] \delta(1-x) \\
   &\quad + \left( C_F - \frac{C_A}{2} \right) \frac{\alpha_s}{\pi}\,\Gamma(1,x) + \mathcal{O}(\alpha_s^2) \,.
\end{aligned}
\end{equation}
Here $\alpha_s\equiv\alpha_s(\mu)$ denotes the renormalized (running) QCD coupling, and 
\begin{equation}
   \Gamma(y,x) = \left[ \frac{\theta(y-x)}{y(y-x)} + \frac{\theta(x-y)}{x(x-y)} \right]_+
\end{equation}
is the symmetric Lange--Neubert kernel \cite{Lange:2003ff}. The plus distribution is defined such that, when $\Gamma(y,x)$ is integrated with a function $f(x)$, one has to replace $f(x)\to f(x)-f(y)$ under the integral. The one-loop coefficient $\gamma_{g,0}'$ vanishes, but we will keep it for later convenience. To all orders in perturbation theory, the coefficient of the scale-dependent logarithm in the anomalous dimension is the difference of the light-like cusp anomalous dimensions in the fundamental and the adjoint representation,
\begin{equation}\label{DeltaGamma}
   \Gamma_F^{\rm cusp}(\alpha_s) - \Gamma_A^{\rm cusp}(\alpha_s) 
   = (C_F-C_A)\,\frac{\alpha_s}{\pi} + \mathcal{O}(\alpha_s^2) \,.
\end{equation}
In the limit $C_A\to 0$, the anomalous dimension in (\ref{eq:gammaJg}) reduces to the anomalous dimension of the radiative jet function with an external photon. We have also derived the anomalous dimension (\ref{eq:gammaJg}) in an independent way using a method recently developed in \cite{Bodwin:2021cpx} (and applied in \cite{Bodwin:2021epw} to the case of the jet function with an external photon). The result we obtain is in agreement with the expression above. Details of this calculation will be presented elsewhere. Using a general relation between an anomalous dimension and the corresponding $Z$-factor derived in \cite{Becher:2009qa}, we find that 
\begin{equation}\label{eq:ZJNLO}
\begin{aligned}
   Z_{J_g}(p^2,x p^2;\mu)
   &= \left[ 1 + (C_F-C_A) \,\frac{\alpha_s}{2\pi} \left( - \frac{1}{\epsilon^{2}}
    + \frac{1}{\epsilon}\,\ln\frac{-p^2-i0}{\mu^2} \right) 
    - \frac{\gamma_{g,0}'\,\alpha_s}{8\pi\epsilon} \right] \delta(1-x) \\
   &\quad + \left( C_F - \frac{C_A}{2} \right) \frac{\alpha_s}{2\pi\epsilon}\,\Gamma(1,x)
    + \mathcal{O}(\alpha_s^2) \,.
\end{aligned}
\end{equation}
The two-loop coefficient in this expression is currently still unknown.

In (\ref{eq:Renormalizationdef}) we have renormalized the jet function without accounting for the renormalization of the (bare) coupling $g_s$ pulled out as a prefactor in (\ref{defJg}). Indeed, in our analysis of $h\to\gamma\gamma$ decay in \cite{Liu:2020wbn} and in the analogous treatment of the $gg\to h$ process in \cite{HiggsGluGlu}, we found it convenient to absorb this bare coupling into the definition of the soft function. An alternative scheme relies of the renormalization of the entire matrix element in (\ref{defJg}), such that
\begin{equation}
   \mu^\epsilon\spac g_s(\mu)\,J_g(p^2,\mu) = \int_0^\infty\!dx\,Z_{J_g}(p^2,x p^2;\mu)\,g_{s,0}\spac J_g^{(0)}(x p^2) 
    \qquad \text{(scheme\,2)} \,.
\end{equation}
The two schemes are related by
\begin{equation}\label{relas}
\begin{aligned}
   Z_{J_g}(p^2,x p^2;\mu) \big|_{\rm scheme\,2} 
   &= Z_{\alpha_s}^{1/2}\,Z_{J_g}(p^2,x p^2;\mu) \big|_{\rm scheme\,1} \,, \\
   \gamma_{J_g}(p^2,x p^2;\mu) \big|_{\rm scheme\,2} 
   &= \gamma_{J_g}(p^2,x p^2;\mu) \big|_{\rm scheme\,1} + \frac{\beta(\alpha_s)}{2\alpha_s}\,\delta(1-x) \,,
\end{aligned}
\end{equation}
where $Z_{\alpha_s}$ is the renormalization factor of the gauge coupling defined by
\begin{equation}
   \alpha_{s,0} = \mu^{2\epsilon}\spac Z_{\alpha_s}\spac\alpha_s(\mu) \,; \qquad
   Z_{\alpha_s} = 1 - \beta_0\,\frac{\alpha_s(\mu)}{4\pi\epsilon} + \mathcal{O}(\alpha_s^2) \,,
\end{equation}
and 
\begin{equation}
   \beta(\alpha_s)
   = \frac{d\alpha_s(\mu)}{d\ln\mu}
   = - 2\alpha_s \sum_{n=0}^\infty\,\beta_n \left( \frac{\alpha_s}{4\pi} \right)^n
\end{equation}
is the QCD $\beta$-function. In order to handle both schemes at a time, we define
\begin{equation}\label{gamg0def}
   \gamma_{g,0}' 
   = \left\{ \begin{array}{ll}
    \,0 \,; & \quad\text{scheme\,1} \\
    \beta_0 \,; & \quad\text{scheme\,2}
    \end{array} \right.
\end{equation} 
in (\ref{eq:gammaJg}), where $\beta_0=\frac{11}{3}\spac C_A-\frac{4}{3}\spac T_F\spac n_f$. 

We now proceed to derive the two-loop expression for the renormalized jet function $J_g(p^2\!,\mu)$. To this end, we first eliminate the bare QCD coupling $\alpha_{s,0}$ in (\ref{Jgbare}) in favor of the renormalized coupling. This yields, again with $\alpha_s\equiv\alpha_s(\mu)$, 
\begin{equation}
\begin{aligned}
   J_g^{(0)}(p^2) 
   &= 1 + \frac{\alpha_s}{4\pi} 
    \left[ e^{-\epsilon\spac L_p} J_{g,1}^{(0)} - \frac{\spac\gamma_{g,0}'}{2\epsilon} \right] \\
   &\quad + \left( \frac{\alpha_s}{4\pi} \right)^2 \left[ e^{-2\epsilon L_p} J_{g,2}^{(0)}
    - \frac{\gamma_{g,0}'+2\beta_0}{2\epsilon}\,e^{-\epsilon L_p} J_{g,1}^{(0)} + \delta_g \right]
    + \mathcal{O}(\alpha_s^3) \,,
\end{aligned}
\end{equation}
where $L_p=\ln[(-p^2-i0)/\mu^2]$. The term $\delta_g$ arises from the two-loop renormalization of the bare coupling $g_s$ in (\ref{defJg}) and is absent in scheme\,1. In scheme\,2 it takes the value $\delta_g=\frac{\beta_1}{4\epsilon}-\frac{\beta_0^2}{8\epsilon^2}$. The second term inside the brackets in the second line generates an extra contribution at two-loop order proportional to $(C_F-C_A)\spac\beta_0$, which contains $1/\epsilon^n$ poles as well as terms that remain finite in the limit $\epsilon\to 0$. We then renormalize the bare jet function (now expressed in terms of the renormalized coupling) using relation (\ref{eq:Renormalizationdef}). At two-loop order this yields 
\begin{equation}
\begin{aligned}
   J_g(p^2,\mu) 
   &= J_g^{(0)}(p^2) + \frac{\alpha_s}{4\pi}\,\int_0^\infty\!dx\,Z_{J_g,1}(p^2,x p^2;\mu) \\
   &\quad + \left( \frac{\alpha_s}{4\pi} \right)^2 \bigg[
    \int_0^\infty\!dx\,Z_{J_g,1}(p^2,x p^2;\mu) \left( e^{-\epsilon\spac L_p} J_{g,1}^{(0)}\,x^{-\epsilon} 
    - \frac{\gamma_{g,0}'}{2\epsilon} \right) \\
   &\hspace{2.47cm} + \int_0^\infty\!dx\,Z_{J_g,2}(p^2,x p^2;\mu) \bigg] + \mathcal{O}(\alpha_s^3) \,,
\end{aligned}
\end{equation}
where $Z_{J_g,n}$ denotes the term in the $Z$-factor multiplying $\left(\frac{\alpha_s}{4\pi}\right)^n$. Note that the term in the renormalization factor (\ref{eq:ZJNLO}) involving the Lange--Neubert kernel gives a non-zero contribution only at two-loop order (first integral in the second line), where it acts on the function $x^{-\epsilon}$. The integral over the two-loop renormalization factor (integral in the third line) is unknown, but in the $\overline{\text{MS}}$ subtraction scheme this integral contains $1/\epsilon^n$ pole terms only. We can thus derive its value from the condition that the expression for the renormalized jet function be finite. Choosing scheme\,1 as an example, this condition yields\footnote{The corresponding relation in scheme\,2 can readily be derived using the first relation in (\ref{relas}).} 
\begin{equation}
   \!\!\int_0^\infty\!dx\,Z_{J_g,2}(p^2,x p^2;\mu) 
   = C_F^2\spac k_{FF} + C_F\spac C_A\spac k_{FA} + C_A^2\spac k_{AA} 
    + C_F\spac T_F\spac n_f\spac k_{F\spac n_f}+ C_A\spac T_F\spac n_f\spac k_{A\spac n_f} \,,
\end{equation}
with
\begin{align}
   k_{FF} &= \frac{2}{\epsilon ^4} - \frac{4 L_p}{\epsilon^3} + \frac{2 L_p^2}{\epsilon^2}
    + \frac{1}{\epsilon} \left( \frac{\pi^2}{2} - 2\zeta_3 \right) , \notag\\
   k_{FA} &= - \frac{4}{\epsilon^4} + \frac{1}{\epsilon^3} \left( 8 L_p + \frac{11}{2} \right)
    + \frac{1}{\epsilon^2} \left( - 4 L_p^2 - \frac{11 L_p}{3} - \frac{67}{18} + \frac{\pi^2}{6} \right) \notag\\
   &\quad + \frac{1}{\epsilon} \left[ \left(\frac{67}{9} - \frac{\pi^2}{3} \right) L_p
    - \frac{202}{27} - \frac{5\pi^2}{9} + 4\zeta_3 \right] , \notag\\
   k_{AA} &= \frac{2}{\epsilon^4} - \frac{1}{\epsilon^3} \left( 4 L_p + \frac{11}{2} \right)
    + \frac{1}{\epsilon^2} \left( 2 L_p^2 + \frac{11 L_p}{3} + \frac{67}{18} - \frac{\pi^2}{6} \right) \\
   &\quad + \frac{1}{\epsilon} \left[ - \left( \frac{67}{9} - \frac{\pi^2}{3} \right) L_p 
    + \frac{395}{54} - 2\zeta_3 \right] , \notag\\
   k_{F\spac n_f} &= - \frac{2}{\epsilon^3} + \frac{2}{\epsilon^2} \left( \frac{4 L_p}{3} + \frac{10}{9} \right)
    + \frac{1}{\epsilon} \left( - \frac{20 L_p}{9} + \frac{56}{27} + \frac{\pi^2}{9} \right) , \notag\\
   k_{A\spac n_f} &= \frac{2}{\epsilon^3} - \frac{2}{\epsilon^2} \left( \frac{4 L_p}{3} + \frac{10}{9} \right)
    + \frac{1}{\epsilon} \left( \frac{20 L_p}{9} - \frac{47}{27} \right) . \notag
\end{align}
The terms multiplying powers of $L_p$ in this expression are governed by the cusp anomalous dimension, such that
\begin{equation}
   \int_0^\infty\!dx\,Z_{J_g,2}(p^2,x p^2;\mu)
   \ni L_p^2\,\frac{\left( \Delta\Gamma_0\right)^2}{8\epsilon^2}
    + L_p \left[  - \frac{\left( \Delta\Gamma_0 \right)^2}{4\epsilon^3}
    - \frac{\beta_0\spac\Delta\Gamma_0}{4\epsilon^2} + \frac{\Delta\Gamma_1}{4\epsilon} \right] ,
\end{equation}
where $\Delta\Gamma_{0,1}$ are the one- and two-loop coefficients of the difference in (\ref{DeltaGamma}), i.e.\ 
\cite{Korchemsky:1987wg,Korchemskaya:1992je}
\begin{equation}
   \Delta\Gamma_0 = 4\spac(C_F-C_A) \,, \qquad
   \frac{\Delta\Gamma_1}{\Delta\Gamma_0} 
   = \left( \frac{67}{9} - \frac{\pi^2}{3} \right) C_A - \frac{20}{9}\,T_F\spac n_f \,.
\end{equation}

After all $1/\epsilon^n$ poles have been subtracted, we take the limit $\epsilon\to 0$ to obtain the two-loop expression for the renormalized jet function. The result is
\begin{align}\label{Jgrenormalized}
   J_g(p^2,\mu) 
   &= 1 + \frac{\alpha_s}{4\pi} \left( C_F - C_A \right) \left( L_p^2 - 1 - \frac{\pi^2}{6} \right) \notag\\
   &\quad + \left( \frac{\alpha_s}{4\pi} \right)^2 \Big[ 
    C_F^2\spac\bar K_{FF} + C_F\spac C_A\spac\bar K_{FA} + C_A^2\spac\bar K_{AA} 
    + C_F\spac T_F\spac n_f\spac\bar K_{F\spac n_f}+ C_A\spac T_F\spac n_f\spac\bar K_{A\spac n_f} \Big] \notag\\
   &\quad + \left( \frac{\alpha_s}{4\pi} \right)^2 \gamma_{g,0}'\,(C_F-C_A)
    \left[ \frac{L_p^3}{6} - \left( \frac12 + \frac{\pi^2}{12} \right) L_p + 1 + \frac73\,\zeta_3 \right] 
    + {\cal O}(\alpha_s^3) \,,   
\end{align}
where $\gamma_{g,0}'$ has been given in (\ref{gamg0def}), and
\begin{equation}
\begin{aligned}
   \bar K_{FF} &= \frac{L_p^4}{2} - \left( 1 + \frac{\pi^2}{6} \right) L_p^2
    + \left( \pi^2 + 4\zeta_3 \right) L_p
    + \frac{3}{2} - \frac{\pi^2}{3} - 39\zeta_3 +\frac{119\pi^4}{360} \,, \\
   \bar K_{FA} &= - L_p^4 - \frac{11 L_p^3}{9} + \frac{85 L_p^2}{9} 
    - \left( \frac{305}{27} + \frac{\pi^2}{2} + 4\zeta_3 \right) L_p
    - \frac{317}{162} - \frac{65\pi^2}{54} + \frac{793\zeta_3}{18} - \frac{143\pi^4}{360} \,,  \\
   \bar K_{AA} &= \frac{L_p^4}{2} + \frac{11 L_p^3}{9} - \left( \frac{76}{9} - \frac{\pi^2}{6} \right) L_p^2
    + \left( \frac{296}{27} - \frac{11\pi^2}{18} \right) L_p
    + \frac{154}{81} + \frac{85\pi^2}{54} - \frac{49\zeta_3}{18} + \frac{\pi^4}{15} \,,  \\
   \bar K_{Fn_f} &= \frac{4 L_p^3}{9} - \frac{20 L_p^2}{9} + \frac{76 L_p}{27} + \frac{14}{81}
    + \frac{5\pi^2}{27} + \frac{8\zeta_3}{9} \,,  \\
   \bar K_{An_f} &= - \frac{4 L_p^3}{9} + \frac{20 L_p^2}{9} 
    - \left( \frac{58}{27} - \frac{2\pi^2}{9} \right) L_p 
    - \frac{275}{81} - \frac{10\pi^2}{27} - \frac{50\zeta_3}{9} \,.
\end{aligned}
\end{equation}
Expression (\ref{Jgrenormalized}) is the main result of this paper. 

\begin{figure}[t]
\begin{center}
\includegraphics[height=0.36\textwidth]{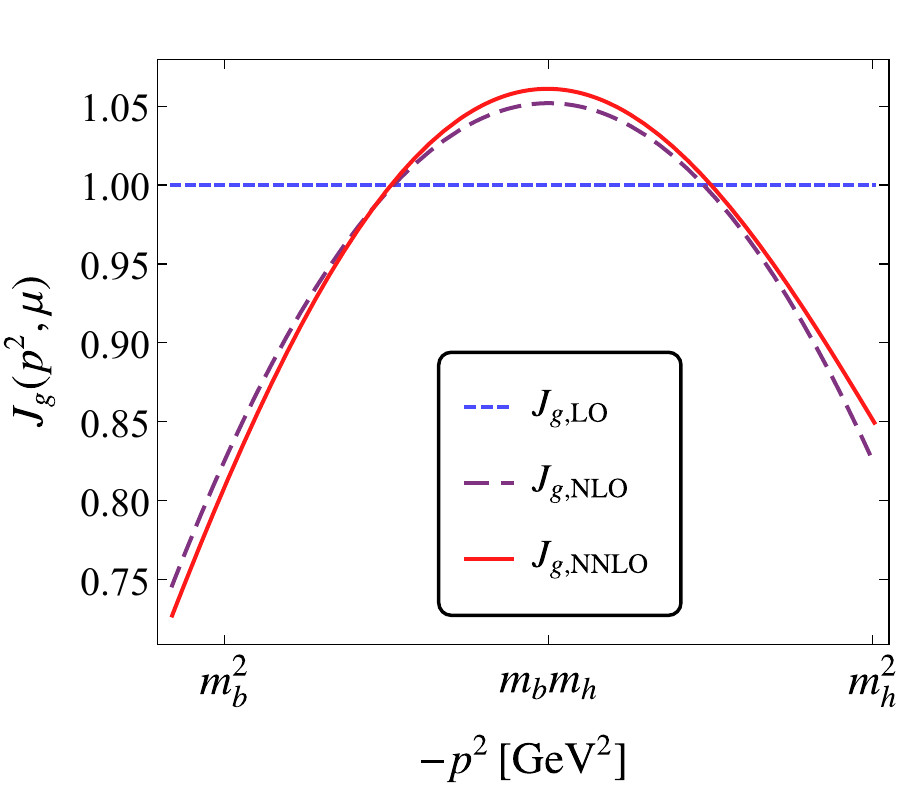} \\
\includegraphics[height=0.36\textwidth]{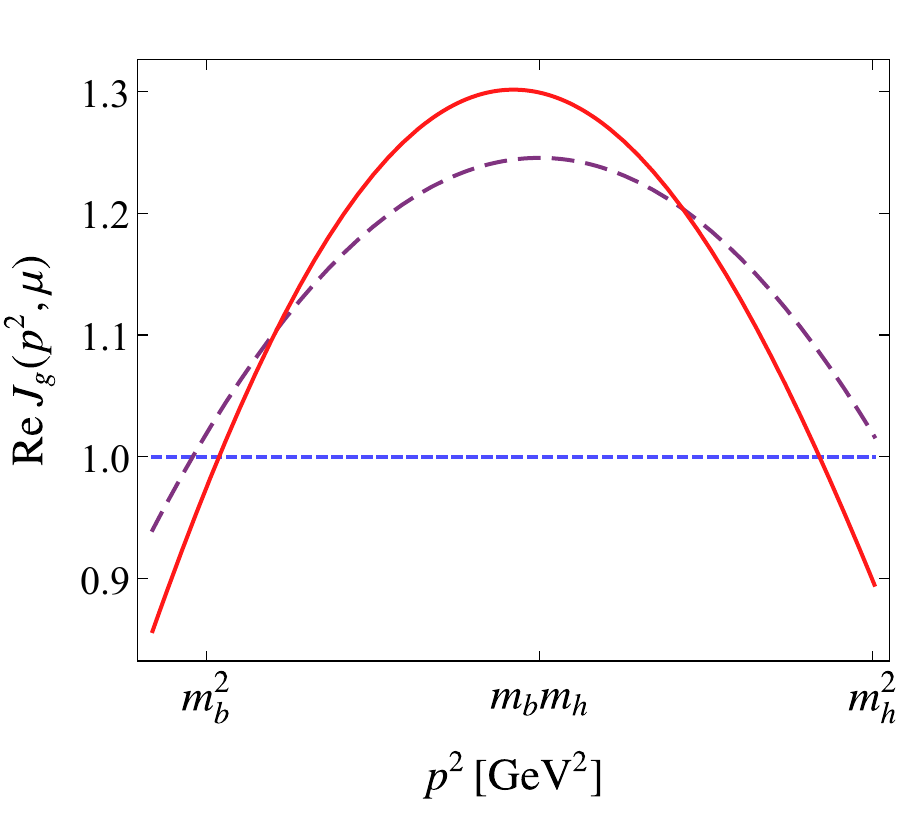} \quad
\includegraphics[height=0.36\textwidth]{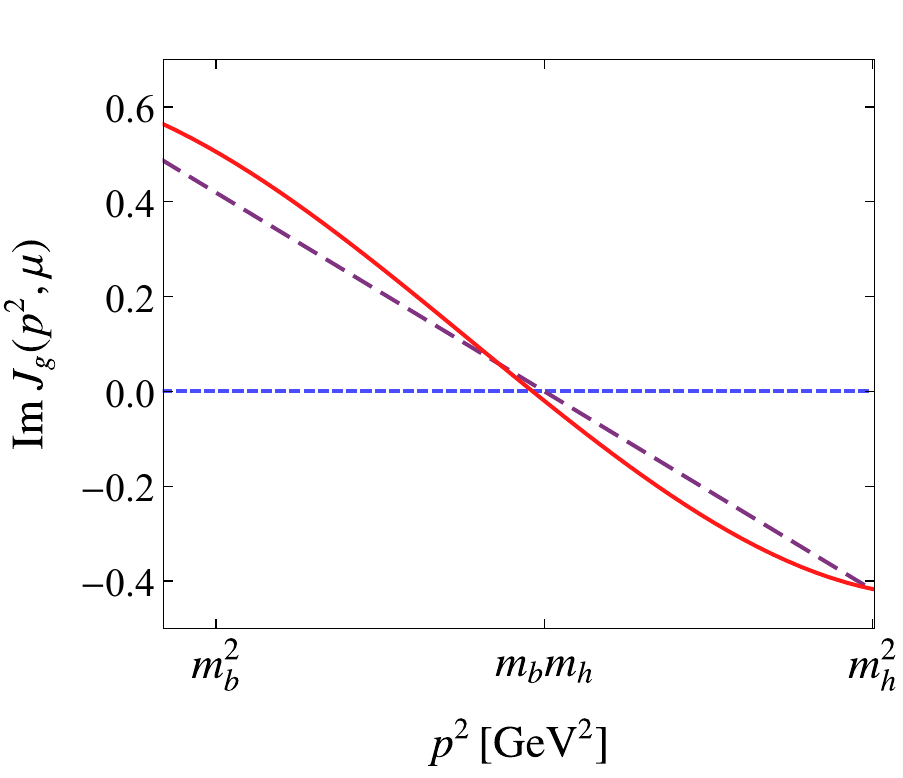}
\caption{\label{fig:Jrenfig} 
Renormalized jet function $J_g(p^2,\mu)$ in the space-like region $p^2<0$ (top) and in the time-like region $p^2>0$ (bottom), evaluated at tree level (dotted) and at one-loop (dashed) and two-loop order (solid) in perturbation theory. The renormalization scale is fixed at $\mu=\sqrt{m_b\spac m_h}$. We use a logarithmic scale on the horizontal axis.}
\end{center}
\end{figure}

In Figure~\ref{fig:Jrenfig} we show the renormalized jet function $J_g(p^2,\mu)$ in scheme\,1 (with $\gamma_{g,0}'=0$) as a function of $p^2$, for both space-like and time-like momenta. For $p^2>0$ the jet function has a non-zero imaginary part, which is shown in the lower-right plot. In the discussion of the Higgs-boson production amplitude $gg\to h$ via $b$-quark loops, the jet function is evaluated for $p^2$ values (of either sign) ranging between $m_b^2$ and $m_h^2$ \cite{HiggsGluGlu}, and this is therefore the range covered in the plots. The renormalization scale is fixed at the geometric mean $\mu=\sqrt{m_b\spac m_h}$. We observe a good convergence of the perturbative expansion, as indicated by the fact that the two-loop contribution to the jet function is considerable smaller than the one-loop contribution for most values of $p^2$. Note that near the boundaries of the plots the logarithm $L_p=\ln(-p^2/\mu^2)$ takes parametrically large values of order $\ln(m_h/m_b)$. In order to obtain reliable results in these regions, these logarithms should be resummed to all orders of perturbation theory, as we will now discuss.

\section{RG evolution of the jet function}

For practical applications in the context of QCD factorization theorems, it is often necessary to evolve a jet function to a scale $\mu$ that is parametrically different from its ``natural'' scale $\mu_j\sim\sqrt{|p^2|}$. To this end, one needs to solve the RG evolution equation (\ref{eq:rge}), which can be accomplished by transforming this equation to Laplace space \cite{Bosch:2003fc,Lange:2003ff,Galda:2020epp}. Here we briefly discuss the solution at leading order in RG-improved perturbation theory, which can be obtained by starting from the solution for the jet function with an external photon derived in \cite{Liu:2020ydl} and accounting for the different color factors. We find
\begin{equation}\label{RGEsol}
\begin{aligned}
   J_g^{\text{RGi,LO}}(p^2,\mu)
   &= e^{-2S_g(\mu_j,\mu)-a_{\gamma_g'}(\mu_j,\mu)}\,\J_g^{\text{LO}}(\partial_\eta,\mu_j)
    \left( \frac{-p^2-i0}{\mu_j^2} \right)^{a_\Gamma(\mu_j,\mu)+\eta} \\
   &\quad\times
    \left[ e^{-2\gamma_E\spac a_\Gamma(\mu_j,\mu)}\,
    \frac{\Gamma\big(1-a_\Gamma(\mu_j,\mu)-\eta\big)\,\Gamma(1+\eta)}%
         {\Gamma\big(1+a_\Gamma(\mu_j,\mu)+\eta\big)\,\Gamma(1-\eta)} \right]^{\rho_g} \Bigg|_{\eta=0} \,, 
\end{aligned}
\end{equation}
where both the Sudakov exponent
\begin{equation}
   S_g(\mu_j,\mu) = \frac{\Delta\Gamma_0}{4\beta_0^2}
    \left[ \frac{4\pi}{\alpha_s(\mu_j)} \left( 1 - \frac{1}{r} - \ln r \right)
    + \left( \frac{\Delta\Gamma_1}{\Delta\Gamma_0} - \frac{\beta_{1}}{\beta_{0}} \right) (1 - r + \ln r)
    + \frac{\beta_1}{2\beta_0} \ln^2 r \right] ,
\end{equation}
with $r=\alpha_s(\mu)/\alpha_s(\mu_j)$, and the function
\begin{equation}\label{eq4.3}
   a_\Gamma(\mu_j,\mu) = \frac{\Delta\Gamma_0}{2\beta_0}\,\ln\frac{\alpha_s(\mu)}{\alpha_s(\mu_j)}
\end{equation}
differ from the corresponding objects in the photon case by the overall color factor $(C_F-C_A)$ (instead of $C_F$). Another important difference is the appearance of the exponent
\begin{equation}
   \rho_g = \frac{C_F-\frac12\spac C_A}{C_F-C_A} = \frac{1}{N_c^2+1} \,,
\end{equation}
which strongly suppresses the deviation of the term shown in the second line of (\ref{RGEsol}) from 1. Note that a consistent evaluation of the Sudakov exponent requires the two-loop coefficients of the cusp anomalous dimension and the QCD $\beta$-function. In analogy with (\ref{eq4.3}), we define
\begin{equation}
   a_{\gamma_g'}(\mu_j,\mu) = \frac{\gamma_{g,0}'}{2\beta_0}\,\ln\frac{\alpha_s(\mu)}{\alpha_s(\mu_j)}
   = \left\{ \begin{array}{ll}
    \hspace{7mm} 0 \,; & \quad\text{scheme\,1} \,, \\
    \ln\!\left[ \frac{\alpha_s(\mu)}{\alpha_s(\mu_j)} \right]^{1/2} ; & \quad\text{scheme\,2} \,.
    \end{array} \right.
\end{equation}
Note that in scheme\,2 the factor
\begin{equation}
   e^{-a_{\gamma_g'}(\mu_j,\mu)}
   = \left[ \frac{\alpha_s(\mu_j)}{\alpha_s(\mu)} \right]^{1/2} 
    = \frac{g_s(\mu_j)}{g_s(\mu)} 
\end{equation}
precisely compensates the scale dependence of $g_s(\mu)$ in the product $g_s(\mu)\,J_g(p^2,\mu)$, as it must be the case.

\begin{figure}[t]
\begin{center}
\includegraphics[height=0.36\textwidth]{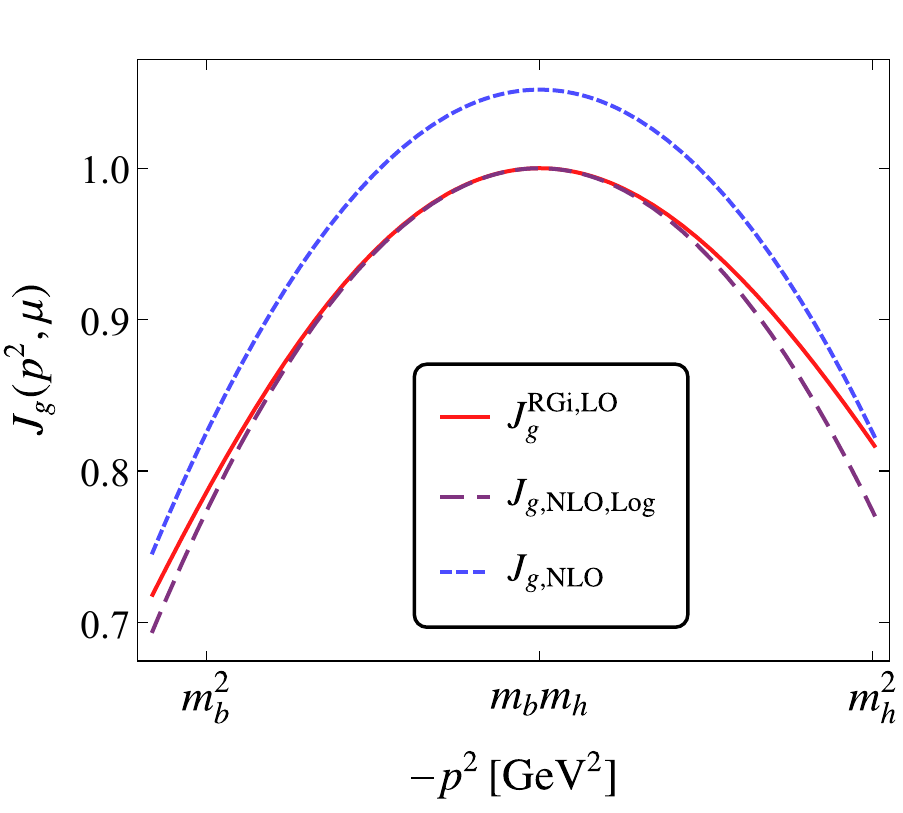} \\
\includegraphics[height=0.36\textwidth]{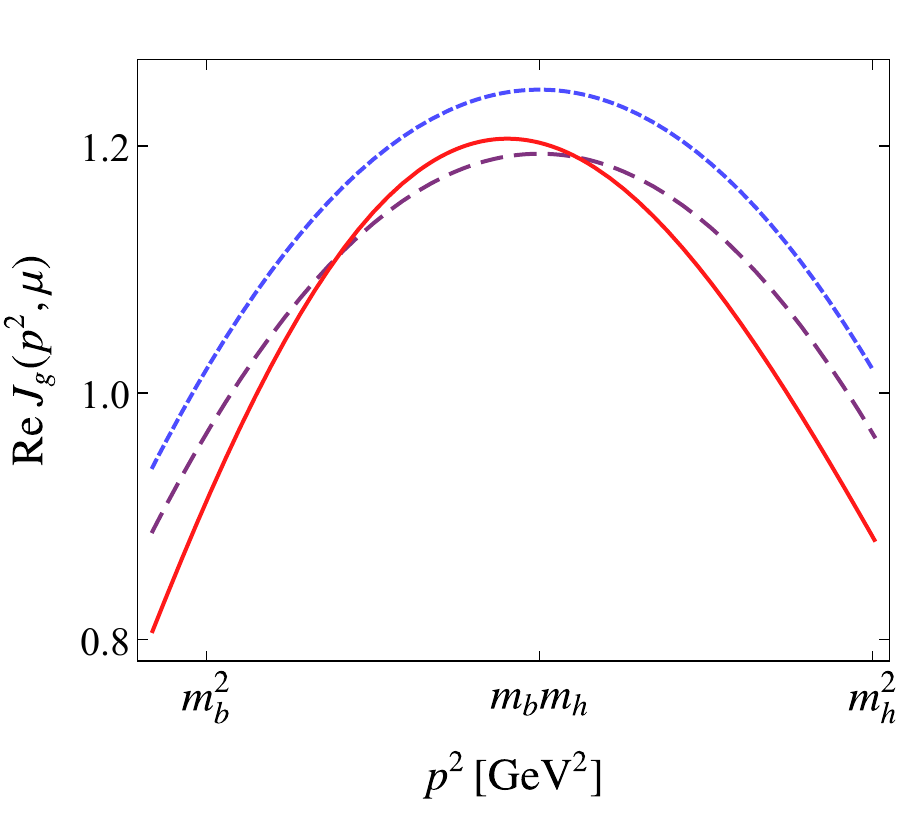} \quad
\includegraphics[height=0.36\textwidth]{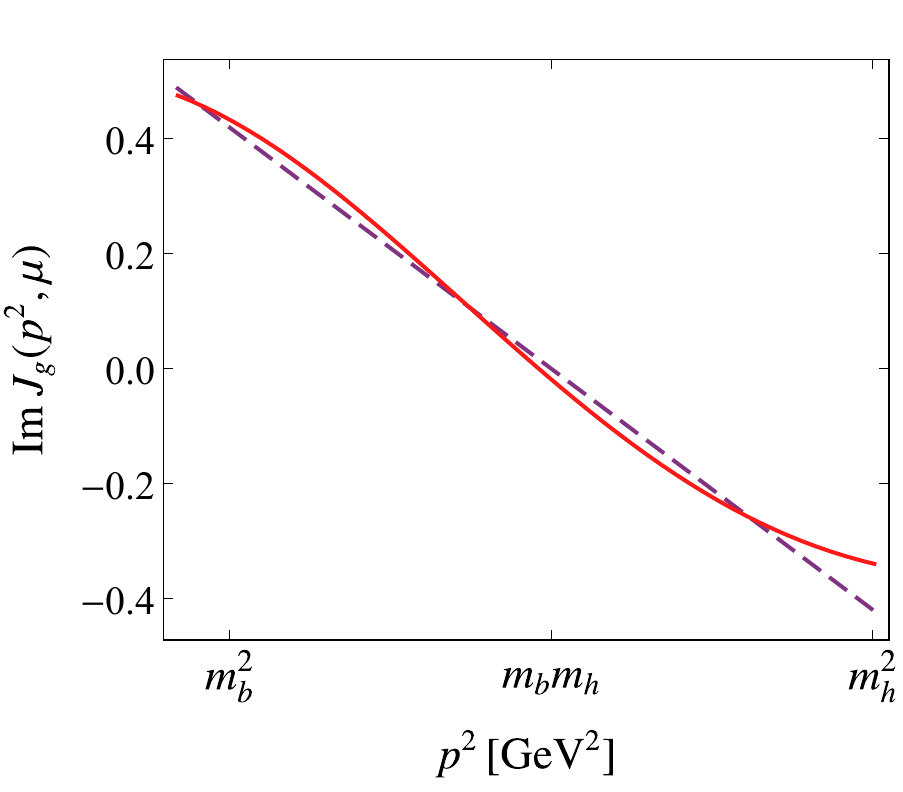}
\caption{\label{fig:JRGEfig} 
Renormalized jet function $J_g(p^2,\mu)$ in the space-like region $p^2<0$ (top) and in the time-like region $p^2>0$ (bottom), at leading order in RG-improved perturbation theory (solid lines). The renormalization scale is fixed to $\mu=\sqrt{m_b\spac m_h}$, while the matching scale $\mu_j$ is chosen as $\mu_j^2=-p^2-i0$. For comparison we also show the one-loop expression for the jet function (dotted lines) and the logarithmic terms in the one-loop expression (dashed lines). In the lower right plot these two curves overlap. We use a logarithmic scale on the horizontal axis.}
\end{center}
\end{figure}

The function $\J_g^{\text{LO}}(\partial_\eta,\mu_j)$ in the solution (\ref{RGEsol}) is a differential operator acting on functions of the auxiliary parameter $\eta$. It is defined by the identification $J_g(p^2,\mu_j)\equiv \J_g(L_p,\mu_j)$. At the matching scale $\mu_j$ chosen as $\mu_j^2\sim p^2$, the expression for the jet function in (\ref{Jgrenormalized}) is free of large logarithms. In fact, at leading order in RG-improved perturbation theory one has $\J_g^{\text{LO}}(L_p,\mu_j)=1$. Using this initial condition, we find the simple result
\begin{equation}
   J_g^{\text{RGi,LO}}(p^2,\mu)
   = e^{-2S_g(\mu_j,\mu)-a_{\gamma_g'}(\mu_j,\mu)} \left( \frac{-p^2-i0}{\mu_j^2} \right)^{a_\Gamma(\mu_j,\mu)} 
    \left[ e^{-2\gamma_E\spac a_\Gamma(\mu_j,\mu)}\,
    \frac{\Gamma\big(1-a_\Gamma(\mu_j,\mu)\big)}{\Gamma\big(1+a_\Gamma(\mu_j,\mu)\big)} \right]^{\rho_g} .
\end{equation}
This formula resums the leading logarithmic corrections to the jet function to all orders of perturbation theory.

In Figure~\ref{fig:JRGEfig} we show the resummed jet function $J_g(p^2,\mu)$ at leading-order in RG-improved perturbation theory (solid lines), for both space-like and time-like momentum transfer. As in Figure~\ref{fig:Jrenfig} we fix the renormalization scale at $\mu=\sqrt{m_b\spac m_h}$. The matching scale $\mu_j$ is chosen as $\mu_j^2=-p^2$ so as to eliminate large logarithms from the matching condition. For comparison, the dotted lines show the one-loop result for the jet function in fixed-order perturbation theory, while the dashed lines show the contribution of only the logarithmic terms in the one-loop result. Comparing the solid and dashed curves, we see that the resummation of higher-order logarithmic effects is important if $|p^2|$ is much larger or much smaller than $\mu^2=m_b\spac m_h$, which is indeed what one would expect. Comparing the dotted and dashed curves shows that the non-logarithmic contributions in the jet function are however also significant. In order to derive a more accurate result at next-to-leading order in RG-improved perturbation theory, it would be necessary to calculate the two-loop contribution to the anomalous dimension in (\ref{eq:gammaJg}). This is an interesting target for future research.

\section{Conclusions} 

We have presented a detailed study of the quark jet function with a radiated gluon, $J_g(p^2,\mu)$ in \eqref{defJg}, which plays a central role in the factorization theorem for the light-quark induced contribution to the $gg\to h$ production amplitude of the Higgs boson at hadron colliders. We have calculated this jet function up to two-loop order, derived its one-loop anomalous dimension, and solved its RG evolution equation at leading order in RG-improved perturbation theory. We find that the quark jet function with a radiated gluon exhibits a more interesting color structure than the corresponding jet function with a radiated photon, considered in \cite{Liu:2020ydl}. While the one-loop gluon jet function can be obtained from the photon jet function by the simple replacement $C_F\to(C_F-C_A)$ of color factors, the one-loop anomalous dimension of the gluon jet function shown in (\ref{eq:gammaJg}) features two different color factors, $(C_F-C_A)$ and $(C_F-\frac{C_A}{2})$. Consequently, the color factor of the Lange--Neubert kernel is no longer the same as that of the Sudakov term, and as a result the leading-order solution to the RG evolution equation in (\ref{RGEsol}) involves the non-trivial exponent $\rho_g=\frac{1}{10}$. Starting at two-loop order, there is no simple relation between the two-loop coefficients of the gluon and photon jet functions.

The results derived in this paper provide an important input to the factorization theorem for the already-mentioned gluon-gluon fusion process $gg\to h$. The quark jet function with an external gluon is a crucial ingredient in two $d$-dimensional refactorization conditions, which allow us to remove the endpoint divergences in the factorization theorem using suitable subtractions \cite{Liu:2019oav,Liu:2020tzd,Liu:2020wbn}. Moreover, the knowledge of the RG equation of the jet function will enable us to derive some logarithmically enhanced contributions to the three-loop $gg\to h$ amplitude, which may be relevant for phenomenology \cite{HiggsGluGlu}.

\paragraph{Acknowledgements:}	
The research of M.N., M.S.\ and X.W.\ is supported by the Cluster of Excellence {\em Precision Physics, Fundamental Interactions and Structure of Matter\/} (PRISMA${}^+$, EXC~2118/1) within the German Excellence Strategy (project ID 39083149). The research of Z.L.L.\ is supported by the Swiss National Science Foundation under grant 200020\_182038.

\begin{appendix}

\section{\boldmath Bare two-loop coefficients to ${\cal O}(\epsilon^2)$}
\label{app:A}
\renewcommand{\theequation}{A.\arabic{equation}}
\setcounter{equation}{0}

In (\ref{eq:2.6}) and (\ref{eq:2.7}), the expansion coefficients in the two-loop expression for the bare jet function (\ref{Jgbare}) have been shown up to terms of ${\cal O}(\epsilon^0)$ in the Laurent series. However, we have also obtained expressions for these coefficients to higher order in $\epsilon$. If in the future one aims at a calculation of the renormalized jet function at three-loop order, then the coefficients are needed up to terms of ${\cal O}(\epsilon^2)$. We find
\begin{align}
   K_{FF} &= \frac{2}{\epsilon ^4}+\frac{1}{\epsilon ^2}\left(-2-\frac{\pi ^2}{3}\right)
    +\frac{1}{\epsilon }\left(-4-\frac{\pi ^2}{2}-\frac{46 \zeta _3}{3}\right)
    +\left(-\frac{13}{2}-\frac{\pi^2}{6}-39 \zeta_3+\frac{\pi ^4}{5}\right) \notag\\
   &\quad +\left(-\frac{33}{4}+\frac{2 \pi ^2}{3}-\frac{431 \zeta _3}{3}-\frac{11 \pi ^4}{15}
    +\frac{26 \pi ^2 \zeta _3}{9}+\frac{766 \zeta_5}{5}\right) \epsilon \notag\\
   &\quad +\left(-\frac{19}{8}+\frac{37 \pi ^2}{12}-\frac{1780\zeta _3}{3} 
    -\frac{116 \pi ^4}{45} +\frac{89 \pi^2\zeta _3}{6}-153 \zeta _5
    +\frac{3707 \pi ^6}{5670}+\frac{2077 \zeta _3^2}{9}\right)\epsilon ^2 \notag\\
   &\quad +O(\epsilon^3) \,, \\[1mm]
   K_{FA} &= -\frac{4}{\epsilon ^4}+\frac{11}{6 \epsilon ^3}
    +\frac{1}{\epsilon ^2}\left(\frac{139}{18}
    +\frac{\pi ^2}{2}\right)+\frac{1}{\epsilon }\left(\frac{319}{27}
    -\frac{\pi^2}{18}+\frac{80 \zeta _3}{3}\right) \notag\\
   &\quad +\bigg(\frac{1087}{162}-\frac{83 \pi^2}{54}+\frac{485 \zeta _3}{18}-\frac{49 \pi ^4}{360}\bigg) \notag\\
   &\quad +\left(-\frac{42551}{972}-\frac{589 \pi ^2}{162}+\frac{7945\zeta _3}{54}
    +\frac{229 \pi^4}{720}-\frac{9 \pi ^2\zeta _3}{2} -\frac{762 \zeta _5}{5}\right) \epsilon \notag\\
   &\quad +\bigg(-\frac{1581791}{5832}-\frac{6919 \pi^2}{972}+\frac{65527\zeta _3}{81}
    +\frac{4433 \pi ^4}{2160}-\frac{1529 \pi^2\zeta _3}{108} -\frac{427 \zeta _5}{30} \notag\\
   &\hspace{1.23cm} -\frac{17471 \pi ^6}{22680}-\frac{5659 \zeta _3^2}{18}\bigg) \epsilon ^2
    +O(\epsilon^3) \,, \\[1mm]
   K_{AA} &= \frac{2}{\epsilon ^4}-\frac{11}{6 \epsilon ^3}
    +\frac{1}{\epsilon ^2}\left(-\frac{103}{18}-\frac{\pi ^2}{6}\right)
    +\frac{1}{\epsilon}\left(-\frac{413}{54}+\frac{11 \pi^2}{18}-\frac{34 \zeta _3}{3}\right) \notag\\
   &\quad +\bigg(\frac{100}{81}+\frac{47 \pi^2}{27}+\frac{259 \zeta _3}{18}-\frac{23 \pi ^4}{360}\bigg) \notag\\
   &\quad +\left(\frac{14375}{243}+\frac{979 \pi ^2}{324}-\frac{103 \zeta _3}{54}+\frac{1117 \pi^4}{2160}
    +\frac{29 \pi ^2 \zeta _3}{18}-\frac{4 \zeta _5}{5}\right) \epsilon \notag\\
   &\quad +\bigg(\frac{220477}{729}+\frac{935 \pi^2}{243}-\frac{17257\zeta _3}{81}
    +\frac{769 \pi ^4}{1296}-\frac{131 \pi^2\zeta _3}{108} +\frac{5947 \zeta _5}{30}
    +\frac{881 \pi ^6}{7560}  +\frac{1505 \zeta _3^2}{18}\bigg) \epsilon ^2 \notag\\
   &\quad +O(\epsilon^3) \,, \\[1mm]
   K_{F\spac n_f} &= -\frac{2}{3 \epsilon ^3}-\frac{10}{9 \epsilon ^2}
    +\frac{1}{\epsilon }\left(-\frac{20}{27}+\frac{\pi^2}{9}\right)
    +\left(\frac{230}{81}+\frac{5 \pi^2}{27}+\frac{64 \zeta _3}{9}\right) \notag\\
   &\quad +\bigg(\frac{4591}{243}+\frac{10 \pi ^2}{81}
    +\frac{320 \zeta _3}{27}+\frac{19 \pi ^4}{180}\bigg) \epsilon \notag\\
   &\quad +\left(\frac{113863}{1458}-\frac{115 \pi^2}{243}+\frac{640\zeta _3}{81}+\frac{19 \pi ^4}{108}
    -\frac{32 \pi ^2\zeta _3}{27} +\frac{544 \zeta _5}{15}\right) \epsilon ^2 
    + O(\epsilon^3) \,, \\[1mm]
   K_{A\spac n_f} &= \frac{2}{3 \epsilon ^3}+\frac{10}{9 \epsilon ^2}
    +\frac{1}{\epsilon}\left(\frac{11}{27}-\frac{2 \pi ^2}{9}\right)
    +\left(-\frac{491}{81}-\frac{10\pi^2}{27}-\frac{106\zeta_3}{9}\right) \notag\\
   &\quad +\left(-\frac{8839}{243}-\frac{67 \pi^2}{162}-\frac{530 \zeta _3}{27}
    -\frac{167 \pi ^4}{540}\right)\epsilon \notag\\
   &\quad +\left(-\frac{111719}{729}+\frac{271 \pi^2}{486}-\frac{1690 \zeta _3}{81}-\frac{167 \pi ^4}{324}
    +\frac{61 \pi ^2 \zeta _3}{27}-\frac{1474 \zeta _5}{15}\right) \epsilon ^2
    +O(\epsilon^3) \,.
\end{align}

\end{appendix}


\begin{thebibliography}{99}

\bibitem{Bauer:2002nz}
C.~W.~Bauer, S.~Fleming, D.~Pirjol, I.~Z.~Rothstein and I.~W.~Stewart,
Phys. Rev. D \textbf{66} (2002), 014017
[arXiv:hep-ph/0202088 [hep-ph]].

\bibitem{Beneke:2002ph}
M.~Beneke, A.~P.~Chapovsky, M.~Diehl and T.~Feldmann,
Nucl. Phys. B \textbf{643} (2002), 431-476
[arXiv:hep-ph/0206152 [hep-ph]].

\bibitem{Moult:2019mog}
I.~Moult, I.~W.~Stewart and G.~Vita,
JHEP \textbf{11} (2019), 153
[arXiv:1905.07411 [hep-ph]].

\bibitem{Beneke:2019kgv}
M.~Beneke, M.~Garny, R.~Szafron and J.~Wang,
JHEP \textbf{09} (2019), 101
[arXiv:1907.05463 [hep-ph]].

\bibitem{Moult:2019uhz}
I.~Moult, I.~W.~Stewart, G.~Vita and H.~X.~Zhu,
JHEP \textbf{05} (2020), 089
[arXiv:1910.14038 [hep-ph]].

\bibitem{Beneke:2019oqx}
M.~Beneke, A.~Broggio, S.~Jaskiewicz and L.~Vernazza,
JHEP \textbf{07} (2020), 078
[arXiv:1912.01585 [hep-ph]].

\bibitem{Moult:2019vou}
I.~Moult, G.~Vita and K.~Yan,
JHEP \textbf{07} (2020), 005
[arXiv:1912.02188 [hep-ph]].

\bibitem{Liu:2019oav}
Z.~L.~Liu and M.~Neubert,
JHEP \textbf{04} (2020), 033
[arXiv:1912.08818 [hep-ph]].

\bibitem{Beneke:2020ibj}
M.~Beneke, M.~Garny, S.~Jaskiewicz, R.~Szafron, L.~Vernazza and J.~Wang,
JHEP \textbf{10} (2020), 196
[arXiv:2008.04943 [hep-ph]].

\bibitem{Liu:2020tzd}
Z.~L.~Liu, B.~Mecaj, M.~Neubert and X.~Wang,
Phys. Rev. D \textbf{104} (2021) no.1, 014004
[arXiv:2009.04456 [hep-ph]].

\bibitem{Liu:2020wbn}
Z.~L.~Liu, B.~Mecaj, M.~Neubert and X.~Wang,
JHEP \textbf{01} (2021), 077
[arXiv:2009.06779 [hep-ph]].

\bibitem{Liu:2020ydl}
Z.~L.~Liu and M.~Neubert,
JHEP \textbf{06} (2020), 060
[arXiv:2003.03393 [hep-ph]].

\bibitem{Lunghi:2002ju}
E.~Lunghi, D.~Pirjol and D.~Wyler,
Nucl. Phys. B \textbf{649} (2003), 349-364
[arXiv:hep-ph/0210091 [hep-ph]].

\bibitem{Bosch:2003fc}
S.~W.~Bosch, R.~J.~Hill, B.~O.~Lange and M.~Neubert,
Phys. Rev. D \textbf{67} (2003), 094014
[arXiv:hep-ph/0301123 [hep-ph]].

\bibitem{DelDuca:1990gz}
V.~Del Duca,
Nucl. Phys. B \textbf{345} (1990), 369-388.

\bibitem{Bonocore:2015esa}
D.~Bonocore, E.~Laenen, L.~Magnea, S.~Melville, L.~Vernazza and C.~D.~White,
JHEP \textbf{06} (2015), 008
[arXiv:1503.05156 [hep-ph]].

\bibitem{Bonocore:2016awd}
D.~Bonocore, E.~Laenen, L.~Magnea, L.~Vernazza and C.~D.~White,
JHEP \textbf{12} (2016), 121
[arXiv:1610.06842 [hep-ph]].

\bibitem{HiggsGluGlu}
Z.~Liu, M.~Neubert, M.~Schnubel and X.~Wang,
in preparation.

\bibitem{tHooft:1975uxh}
G.~'t Hooft,
{\em The Background Field Method in Gauge Field Theories}, 
Procs.\ 12$^{th}$ Annual Winter School of Theoretical Physics (Karpacz, Poland, 1975).

\bibitem{Boulware:1980av}
D.~G.~Boulware,
Phys. Rev. D \textbf{23} (1981), 389.

\bibitem{Abbott:1980hw}
L.~F.~Abbott,
Nucl. Phys. B \textbf{185} (1981), 189-203.

\bibitem{Abbott:1981ke}
L.~F.~Abbott,
Acta Phys. Polon. B \textbf{13} (1982), 33.

\bibitem{Meissner:1986tr}
K.~A.~Meissner,
Acta Phys. Polon. B \textbf{17} (1986), 409-416.

\bibitem{Becher:2010pd}
T.~Becher and G.~Bell,
Phys. Lett. B \textbf{695} (2011), 252-258
[arXiv:1008.1936 [hep-ph]].

\bibitem{Lee:2013mka}
R.~N.~Lee,
J. Phys. Conf. Ser. \textbf{523} (2014), 012059
[arXiv:1310.1145 [hep-ph]].

\bibitem{Smirnov:2019qkx}
A.~V.~Smirnov and F.~S.~Chuharev,
Comput. Phys. Commun. \textbf{247 } (2020), 106877
[arXiv:1901.07808 [hep-ph]].

\bibitem{Klappert:2020nbg}
J.~Klappert, F.~Lange, P.~Maierh\"ofer and J.~Usovitsch,
Comput. Phys. Commun. \textbf{266} (2021), 108024
[arXiv:2008.06494 [hep-ph]].

\bibitem{Bosch:2004th}
S.~W.~Bosch, B.~O.~Lange, M.~Neubert and G.~Paz,
Nucl. Phys. B \textbf{699} (2004), 335-386
[arXiv:hep-ph/0402094 [hep-ph]].

\bibitem{Grozin:1996pq}
A.~G.~Grozin and M.~Neubert,
Phys. Rev. D \textbf{55} (1997), 272-290
[arXiv:hep-ph/9607366 [hep-ph]].

\bibitem{Lange:2003ff}
B.~O.~Lange and M.~Neubert,
Phys. Rev. Lett. \textbf{91} (2003), 102001
[arXiv:hep-ph/0303082 [hep-ph]].

\bibitem{Liu:2020eqe}
Z.~L.~Liu, B.~Mecaj, M.~Neubert, X.~Wang and S.~Fleming,
JHEP \textbf{07} (2020), 104
[arXiv:2005.03013 [hep-ph]].

\bibitem{Bodwin:2021cpx}
G.~T.~Bodwin, J.~H.~Ee, J.~Lee and X.~P.~Wang,
Phys. Rev. D \textbf{104} (2021) no.1, 016010
[arXiv:2101.04872 [hep-ph]].

\bibitem{Bodwin:2021epw}
G.~T.~Bodwin, J.~H.~Ee, J.~Lee and X.~P.~Wang,
arXiv:2107.07941 [hep-ph].

\bibitem{Becher:2009qa}
T.~Becher and M.~Neubert,
JHEP \textbf{06} (2009), 081
[Erratum: JHEP \textbf{11} (2013), 024]
[arXiv:0903.1126 [hep-ph]].

\bibitem{Korchemsky:1987wg}
G.~P.~Korchemsky and A.~V.~Radyushkin,
Nucl. Phys. B \textbf{283} (1987), 342-364.

\bibitem{Korchemskaya:1992je}
I.~A.~Korchemskaya and G.~P.~Korchemsky,
Phys. Lett. B \textbf{287} (1992), 169-175.

\bibitem{Galda:2020epp}
A.~M.~Galda and M.~Neubert,
Phys. Rev. D \textbf{102} (2020), 071501
[arXiv:2006.05428 [hep-ph]].

\end{thebibliography}
\end{document}